\documentclass[%
 reprint,
 amsmath,amssymb,
 aps,
]{revtex4-2}

\usepackage{graphicx}
\usepackage{dcolumn}
\usepackage{bm}
\usepackage{multirow}
\usepackage{color}
\usepackage{changepage}
\usepackage{hyperref}
\usepackage[normalem]{ulem}

\begin{document}

\preprint{APS/123-QED}

\title{Tidal heating as a direct probe of strangeness inside neutron stars}

\author{Suprovo Ghosh}
\email{suprovo@iucaa.in}
\author{Bikram Keshari Pradhan}
\email{bikramp@iucaa.in}
\author{Debarati Chatterjee}%
\email{debarati@iucaa.in}
\affiliation{%
 Inter-University Centre for Astronomy and Astrophysics,
Pune University Campus,
Pune 411007, India\\
}%


\newcommand{\sg}[1]{\textsf{\color{red}{ #1}}}
\newcommand{\dc}[1]{\textsf{\color{red}{ #1}}}
\newcommand{\bkp}[1]{\textsf{\color{green}{ #1}}}

\begin{abstract}
It has been discussed whether viscous processes in neutron star matter during a binary inspiral can damp out the tidal energy induced by the companion and heat up the star. Earlier investigations concluded that this tidal heating is negligible for normal neutron star matter. In this work, we suggest a novel effect of tidal heating involving strange matter in the neutron star interior, that can significantly heat up the star, and is potentially observable by current and future gravitational wave detectors. We propose that this could serve as a direct probe of strangeness in neutron stars.

\end{abstract}

\maketitle


\section{\label{sec:intro} Introduction}
Neutron stars (NS) are unique astrophysical dense objects to study  cold  and  dense  matter  under  extreme  conditions. In the high density environment of the NS interior, strangeness containing   matter e.g. hyperons, kaons or deconfined quarks, can become stable components and have important consequences for the equation of state (EOS), structure and evolution of these compact objects~\cite{GlendenningBook,Lattimer2004,Blaschke2018,Tolos_2020,Annala_2023}. 
Various astrophysical observations e.g., the measurement of high mass NS~\cite{Weissenborn2011,Weissenborn_2012}, radius measurement from low-mass X-ray binaries~\cite{Constanca2013,Fortin2015}, thermal evolution~\cite{Anzuini_2021} and gravitational wave (GW) emission from  hot and rapidly rotating newly born neutron stars due to unstable r-modes ~\cite{Lindblom2002,Debi_2006,Jha2022} as well as binary neutron star merger event GW170817~\cite{Abbott2017,Abbott2018,Abbott2019} by the LIGO–Virgo detectors~\citep{AdvLIGO2015,Abbott2016,AdvVIRGO2014} have been used to constrain the hyperon content inside their core~\cite{Chatterjee_2016,Gomes_2019,blacker2023thermal}. The measurement of tidal deformability from  GW170817~\cite{Abbott2017} along with  recent NICER estimates of mass-radius posteriors from two pulsars, PSR $J0030+0451$~\cite{NICER0030_Miller,NICER0030_Riley} and PSR $J0740+6620$~\cite{NICER0740_Miller,NICER0740_Riley} have significantly constrained the unknown neutron star EOS~\cite{Annala,De,Most,Dietrich2020,Pang_2021,Issac2021}.  \\

During binary neutron star inspiral, tidal interaction due to the companion transfers mechanical energy and angular momentum to the star at the expense of the orbital energy and the damping mechanisms inside the star convert this energy to heat (``tidal heating"). This tidal heating could potentially cause mass ejection in a radiation-driven outflow prior to the merger~\cite{Rees1992} or spin the stars up to corotation before merger~\cite{Kochanek1992,Bildsten1992}. However this heating is found to be negligible for viscosities arising from  nucleonic or leptonic weak processes, because of the long viscous timescale. Lai (1994)~\cite{Lai_1994} found that shear and bulk viscous damping of mode oscillations of normal NS matter could only heat the stellar core to $\sim 3\times 10^7$K. Recent work by Arras et al.~(2019)~\cite{Arras2019} considers heating due to direct Urca reactions during inspiral and concluded that they are heated upto $T \sim 10^8$K, which is one order of magnitude higher than  the estimates by Lai (1994)~\cite{Lai_1994}. \\

Bulk viscosity originating from non-leptonic weak interaction involving strange matter in the neutron star core have been studied extensively in the context of damping of unstable $r-$mode oscillations ~\cite{Jones2001,Lindblom2002,Debi_2006,Dalen2004,Jha2010,Haskell2010,Ofengeim,Alford2021,Jha2022}. Hyperon bulk viscosity is found to be several orders higher than viscosity due to nucleonic Urca ($\approx 10^{8} - 10^{10}$ times the canonical shear viscosity from $e-e$ scattering in the temperature range $10^{6}-10^8$K) and strongly suppress the r-mode instability below $10^{10}$K~\cite{Lindblom2002,Haskell2010}. In this article, for the first time we propose that this tidal heating due to bulk viscosity from strange matter in the neutron star core can sufficiently heat up the star during the inspiral to temperatures much higher than estimated from earlier studies. This effect is potentially detectable by current and future generation GW detectors as a deviation of the orbital decay rate from the general relativistic point-mass result. \\

The structure of the article is as follows: in Sec.~\ref{sec:HyperonBV}, we compute the hyperon bulk viscosity for several choice of EOS models and compare the strength with canonical shear and bulk viscosity sources inside neutron star cores. In Sec.~\ref{sec:TH_NS}, we describe the formalism to calculate tidal heating in binary neutron star inspiral and estimate the temperatures reached during inspiral due to dissipation from hyperon bulk viscosity. Then in Sec.\ref{sec:phase} , we estimate the phase difference introduced in the gravitational wave signal from binary NS inspiral due to this viscous dissipation and argue that it will be detectable in future generation GW detectors. In Sec.~\ref{sec:discussion} we discuss the main implications of this work and also future directions.

\section{\label{sec:HyperonBV} Hyperonic Bulk viscosity}

\begin{figure}
    \centering
    \includegraphics[width=0.5\textwidth]{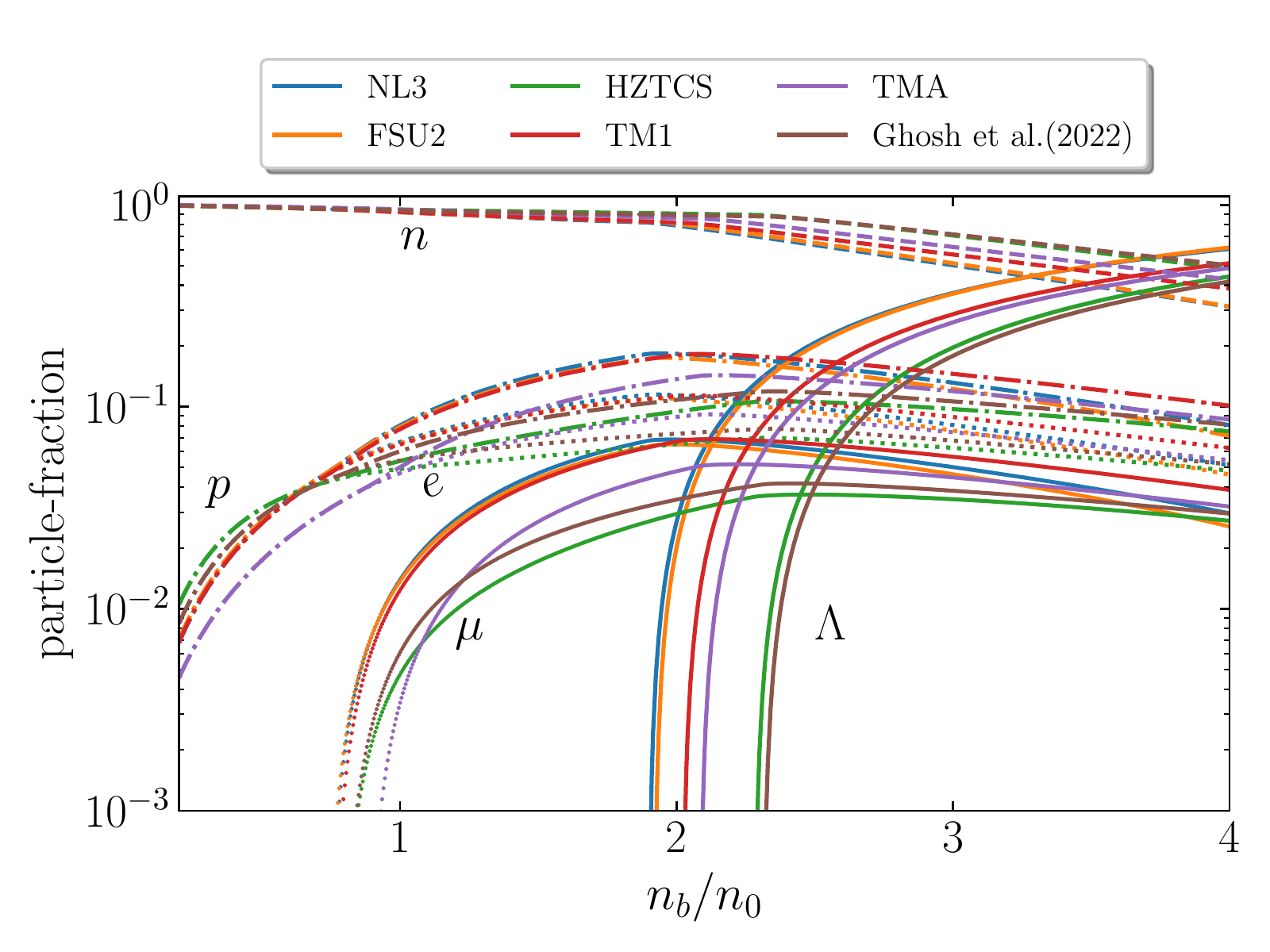}
    \caption{Particle fraction as a function of normalised density($n_0$ is the nuclear saturation density) for the chosen EOS parametrizations given in Table~\ref{tab:table1}. The compositions considered here are : $n$(dashed lines),$p$(dashed dotted line), $e$(dotted lines),$\mu$(dense dotted lines) and $\Lambda$-hyperon(solid lines).}
    \label{fig:fraction}
\end{figure}
\begin{figure}
    \centering
    \includegraphics[width=0.5\textwidth]{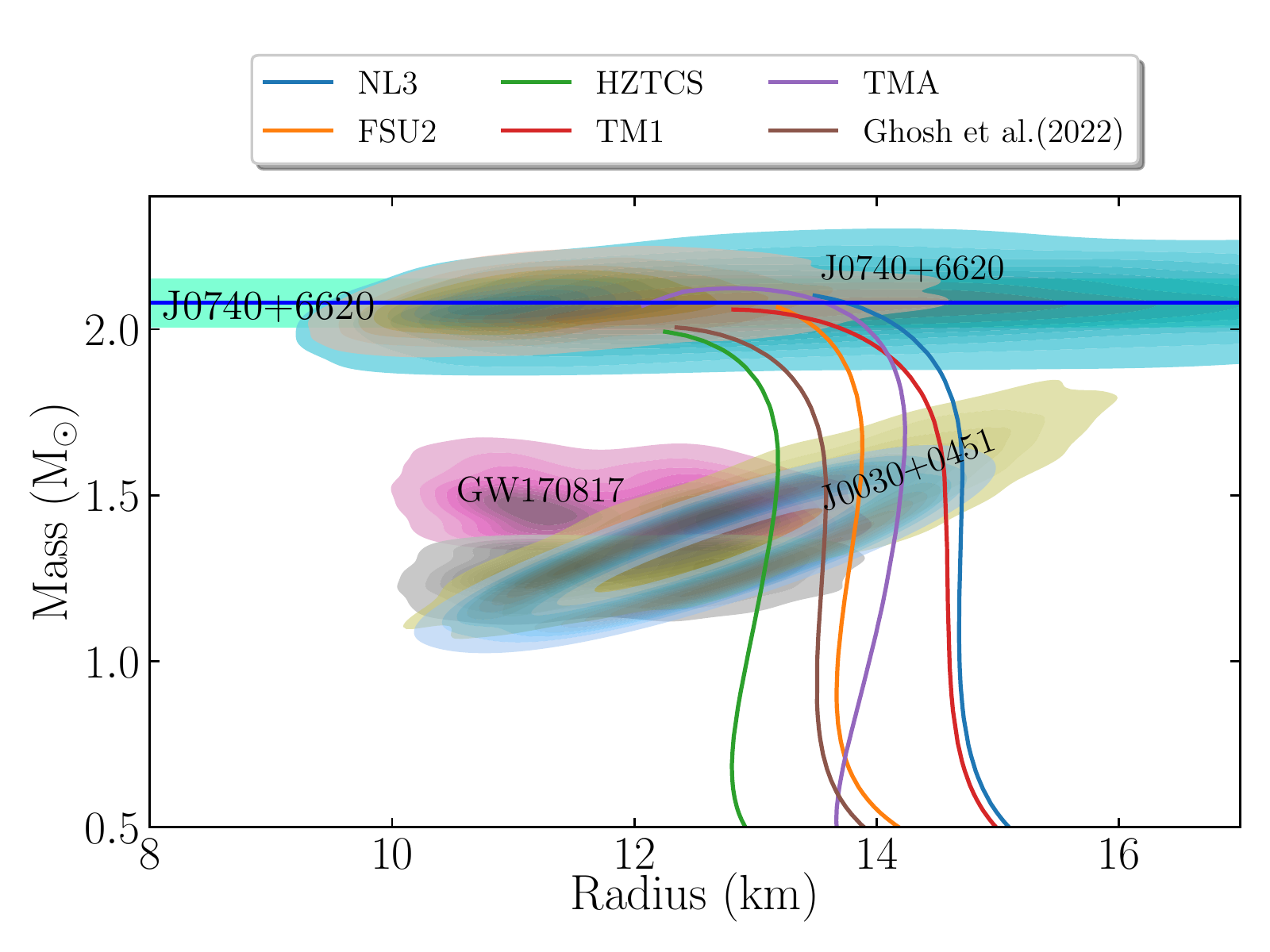}
    \caption{Mass radius diagram for the  different EOSs given in Table~\ref{tab:table1} along with the Mass-Radius posterior from GW170817~\cite{Abbott2017} and $2-\sigma$ NICER posterior of PSR $J0030+0451$~\cite{NICER0030_Miller,NICER0030_Riley} and PSR $J0740+6620$~\cite{NICER0740_Miller,NICER0740_Riley}. The $1-\sigma$ mass estimate of massive pulsar PSR $J0740+6620$~\cite{Fonseca_2021} is shown in light green band with the median value of $2.08M_{\odot}$ shown in the deep blue line.}
    \label{fig:MR}
\end{figure}

In this work, we explore the effect by calculating the bulk viscosity from hyperons within a simple model. During binary inspiral, bulk viscosity arises due to the pressure and density variations associated with the dynamical tidal excitation of NS oscillation modes (e.g. $f$ or $g$-modes) that drive the system out of chemical equilibrium~\cite{Jones2001}. Weak reactions among the core constituents, having timescales comparable to the oscillation modes, bring the system back to equilibrium. Among the possible hyperons that may appear in the core, $\Lambda$ hyperons contribute dominantly to the bulk viscosity~\cite{Jones2001,Lindblom2002}. Nuclear experiments indicate a repulsive $\Sigma$-nucleon potential~\cite{FriedmanGal07,SchaffnerGal00,Mares951} that result in suppression of $\Sigma$-hyperons, while the $\Xi$-hyperons may appear at large densities~\cite{Weissenborn2011} beyond the maximum densities reached even in very massive NSs. However, one must note that for certain parametrizations e.g. FSU2H~\cite{Tolos_2017} and density dependent RMF parametrizations~\cite{Adriana_2020,Tu_2022}, the onset densities of $\Lambda$, $\Sigma$ and $\Xi$ hyperons are quite close. As a proof-of-principle, in this study we consider the dominant non-leptonic weak process involving only the $\Lambda$ hyperon 
\begin{equation}\label{eq:Hyp2}
    \centering
    n + p \longleftrightarrow p + \Lambda .
\end{equation}
Though there may be other processes involving the $\Lambda$ hyperon and also other hyperon species ($\Sigma$ and $\Xi$ hyperons) in NS core~\cite{Gusakov_2008}, this estimate provides a reasonable lower limit on the rates and hence an upper limit on the hyperon bulk viscosity~\cite{Gusakov_2008,Jones2001}. To investigate the effect of the EOS, we consider a diverse set of parametrizations within the Relativistic Mean Field (RMF) model~\cite{GM91,Chen2014,Ghosh2022} consistent with recent multi-messenger observations. The potential depth of $\Lambda$ hyperons in normal nuclear matter is set to $U_{\Lambda}^N = -30 $ MeV~\cite{Weissenborn_2012} for all the parametrizations. In Fig.~\ref{fig:fraction}, we plot the particle fraction of all the constituents($npe\mu$ \& $\Lambda$) for chosen set of RMF parametrizations and also mention the density for appearance of $\Lambda$ hyperon in Table~\ref{tab:table1}. We see that for all the parametrizations, $\Lambda$-hyperon appear first around twice the nuclear saturation density($\sim 2n_0$) and their fraction increase with increasing density.
For the chosen EOSs, we also  plot the mass-radius in Fig.~\ref{fig:MR} along with the astrophysical constraints from binary NS merger event GW170817~\cite{Abbott2017} and NICER estimates of PSR $J0030+0451$~\cite{NICER0030_Miller,NICER0030_Riley} and PSR $J0740+6620$~\cite{NICER0740_Miller,NICER0740_Riley}. The maximum mass reached by all the EOSs are $\geq 2M_{\odot}$ as also mentioned in Table~\ref{tab:table1} and consistent with the observed massive pulsar PSR $J0740+6620$~\cite{Fonseca_2021}. In our study, to consider significant hyperon fraction we choose a couple of stiff EOSs (e.g. NL3, TM1) for which radius just extends($\sim 1$ km) beyond the 90\% confidence interval of GW170817 estimates but they are still well within the limits of NICER estimates of PSR $J0030+0451$. 
\\

\begin{figure}
    \centering
    \includegraphics[width=0.5\textwidth]{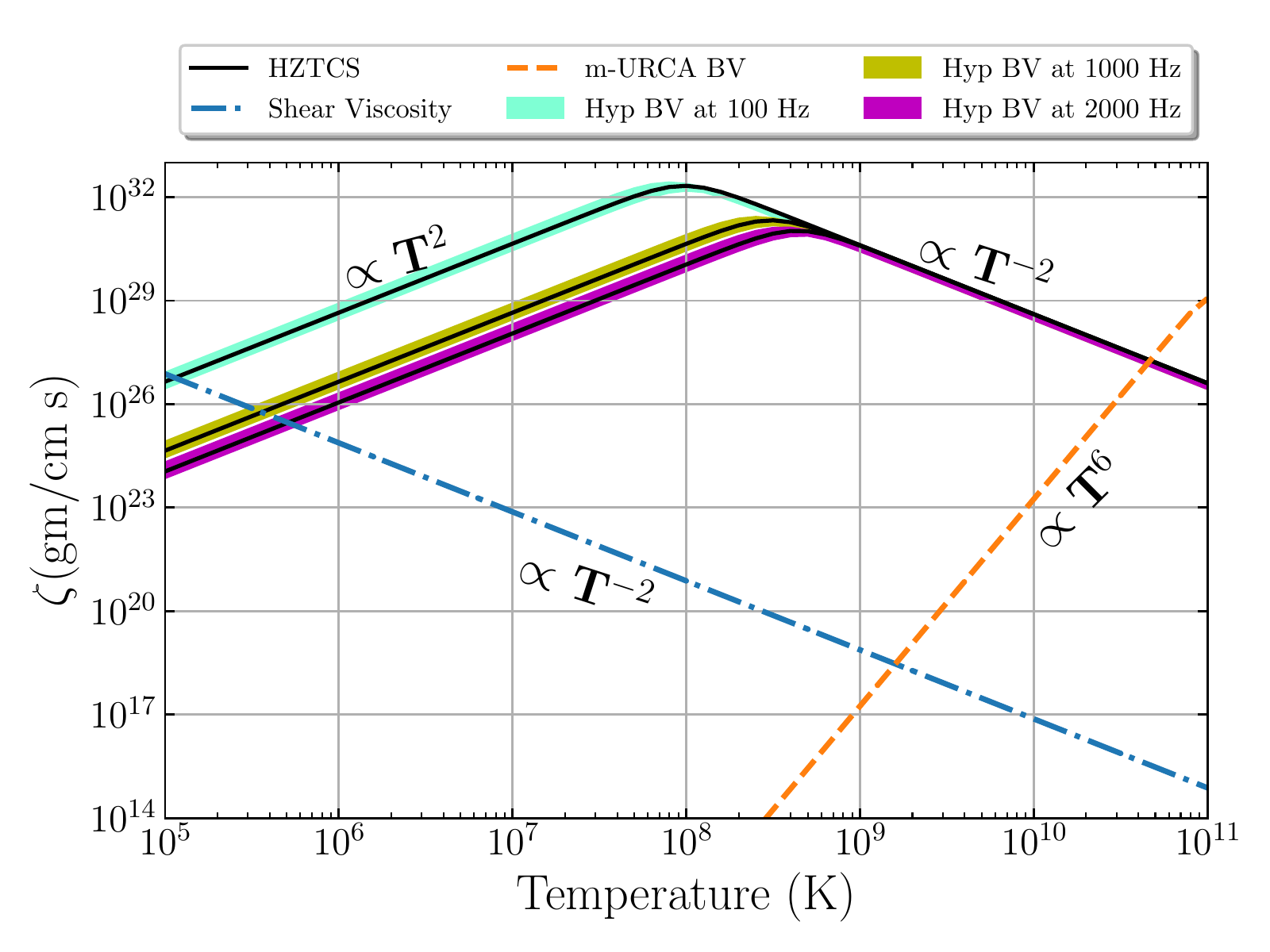}
    \caption{Relative strengths of various sources of viscosities inside NS matter as a function of temperature for $n = 2.5n_0$( $n_0$ = the nuclear saturation density). The bands corresponds to hyperon bulk viscosity coming from  different EOSs given in Table~\ref{tab:table1} with the black solid line corresponding to HZTCS EOS. The standard neutron matter shear viscosity coming from $e-e$ scattering and bulk viscosity from m-Urca reactions at f = $1$kHz are also plotted along with their temperature dependence.}
    \label{fig:fig_1}
\end{figure}

Now, we calculate the hyperon bulk viscosity analytically for these chosen EOSs, whose details are provided in Table~\ref{tab:table1} along with hyperon potentials and onset densities. Given an EOS, the real part of the bulk viscosity coefficient ($\zeta$) can be calculated in terms of relaxation times ($\tau$) of weak processes~\cite{Lindblom2002,Debi_2006}
\begin{equation}\label{BV}
    \zeta = n_B\frac{\partial P}{\partial n_n} \frac{dn_n}{d n_{B}}\frac{\tau}{1+(\omega\tau)^2},
\end{equation}
where  $P$ is the  pressure, $n_B$ the total baryon number density, $n_n$ the neutron density and $\omega$ the frequency of the oscillation mode. The rates $1/\tau$ of these reactions are calculated from tree-level Feynman diagrams involving exchange of a $W$-boson~\cite{Jones2001,Ofengeim,Lindblom2002}. 
In Fig.~\ref{fig:fig_1}, we display the bulk viscosity due to the process~\ref{eq:Hyp2} as a function of temperature for the equations of state in Table~\ref{tab:table1}, for typical characteristic values of mode frequency: $1$ \& $2$ kHz for $f$-mode~\cite{Pradhan_2021} and $\sim 100$ Hz for $g$-mode oscillations~\cite{tran2022gmode}. Depending on the mode oscillation frequency, we see that the resonances occurs in the temperature range of $10^8 - 10^9$K and the resonant maximum value of $\zeta$ can reach values of $10^{31} - 10^{32}$ gm cm$^{-1}$s$^{-1}$, which are several orders of magnitude higher than the canonical shear and bulk viscosity~\cite{Sawyer1989} in this temperature regime. 

\begin{table*}
\begin{adjustwidth}{-3em}{}
\begin{center}
    \begin{tabular}{|c|c|c|c|c|c|c|c|c|c|}
    \hline
    EOS & Max. mass & $\Lambda$ onset & Mass~($M_{\odot}$) & Central  & Radius & Hyperon core& f-mode  & Max. temp (K)  & $\Delta \Phi = 2\pi \Delta\mathcal{N}$  \\ 
    & ($M_{\odot}$) & density & & density & (km) &  radius(km) & Freq.(Hz) &  at $D/R = 3$ & (rad) \\
    \hline
    \multicolumn{10}{c}{} \\
    \hline
    NL3 & 2.10 & 1.90$n_0$ & 1.6 & 2.07$n_0$ & 14.7 & 3.42 & 1847 & 9.7$\times 10^8$ & 0.001 \\
    ~\cite{Chen2014}& & & 1.8 & 2.48$n_0$ & 14.6 & 6.19 & 1909 & 3.3$\times 10^9$ & 0.08 \\
    & & & 2.0 & 3.35$n_0$ & 14.2 & 8.10 & 2009 & 6.2$\times 10^9$ & 0.3 \\
    \hline
    TM1 & 2.06 & 2.02$n_0$ & 1.6 & 2.24$n_0$ & 14.55 & 3.16 & 1873 & 8.7$\times 10^8$ & 0.0008 \\
    ~\cite{Steiner_2013}& & & 1.8 & 2.77$n_0$ & 14.37 & 6.18 & 1947 & 3.4$\times 10^9$ & 0.09 \\
    & & & 2.0 & 4.06$n_0$ & 13.6 & 8.15 & 2092 & 6.7$\times 10^9$ & 0.34 \\
    \hline
    TMA & 2.12 & 2.09$n_0$ & 1.8 & 2.54$n_0$ & 14.2 & 5.13 & 1948 & 2.3$\times 10^9$ & 0.02 \\
    ~\cite{Steiner_2013}& & & 2.0 & 3.35$n_0$ & 13.89 & 7.36 & 1909 & 5.1$\times 10^9$ & 0.16\\
    \hline
    HZTCS & 2.00 & 2.28$n_0$ & 1.6 & 2.67$n_0$ & 13.2 & 4.89 & 2108 & 2.3$\times 10^9$ & 0.02 \\
    ~\cite{HZTCS}& & & 1.8 & 3.32$n_0$ & 13.1 & 6.82 & 2171 & 4.7$\times 10^9$ & 0.16 \\
    & & & 2.0 & 5.32$n_0$ & 12.25 & 8.17 & 2305 & 7.9$\times 10^9$ & 0.44 \\
    \hline
    FSU2 & 2.03 & 1.92$n_0$ & 1.6 & 2.22$n_0$ & 14.4 & 4.98 & 1898 & 2.1$\times 10^9$ & 0.03 \\
    ~\cite{Chen2014}& & & 1.8 & 2.72$n_0$ & 14.2 & 7.07 & 1968 & 4.5$\times 10^9$ & 0.19 \\
    & & & 2.0 & 3.82$n_0$ & 13.6 & 8.54 & 2099 & 7.4$\times 10^9$ & 0.47 \\
    \hline 
    Stiffest EOS & 2.01 & 2.31$n_0$ & 1.6 & 2.71$n_0$ & 13.5 & 3.88 & 2047 & 1.4$\times 10^9$ & 0.004 \\
    from Ghosh et al.& & & 1.8 & 3.39$n_0$ & 13.4 & 6.34 & 2119 & 4.1$\times 10^9$ & 0.11 \\
    (2022)~\cite{Ghosh_2022}& & & 2.0 & 5.5$n_0$ & 12.5 & 8.05 & 2256 & 7.2$\times 10^9$ & 0.37 \\
    \hline
    \end{tabular} 
\end{center}
\end{adjustwidth}
    \caption{Detailed list of NS properties corresponding to different EOS parametrizations considered in this work. The potential depth of $\Lambda$ hyperons in normal nuclear matter is set to $U_{\Lambda}^N = -30 $ MeV~\cite{Weissenborn_2012} for all the parametrizations. Densities are given in terms of $n_0$, the nuclear saturation density ($\sim 0.15 $ fm$^{-3}$~\cite{Oertel2017}). The phase difference $\Delta\Phi$ in the last column denotes the total phase difference accumulated at the end of inspiral around $f \sim 500$Hz.}
    \label{tab:table1}
\end{table*}

\section{Tidal Heating in binary NS}
\label{sec:TH_NS}
To investigate the effect of this hyperon viscous damping on heating of the NS, we first analyse the timescale for mode damping. The mode damping rate is given by
\begin{equation}\label{eq:damptime2}
    \gamma_{visc,\alpha} = \frac{1}{2E_{\alpha}}\Dot{E}_{visc,\alpha}
\end{equation}
where $\alpha$ denotes the particular eigenmode, $E_{\alpha}$ is the energy of the mode and $\Dot{E}_{visc,\alpha}$ is the energy dissipation rate. In the adiabatic approximation, the effect of the tidal potential due to the companion star is measured in terms of the Lagrangian fluid displacement vector $\bm{\xi}(r,t)$ from its equilibrium position. This displacement can be analysed in terms of the normal modes of the NS.  In this work, we  only consider the dominant $ l = m = 2$ fundamental ($f)$ mode contribution to the tidal heating. We determine the $f$-mode frequency via the Cowling approximation~\cite{Pradhan_2021} and the normalised mode eigenfunctions are used to calculate the tidal coupling~\cite{Andersson_2018}. Although general relativistic effects can modify $f$-mode frequencies by up to $30\%$~\cite{Pradhan_fullGR}, we restrict ourselves to Cowling approximation to be consistent with the Newtonian description of  tidal heating used in this work. The bulk viscous dissipation rate is expressed as~\cite{Lai_1994} 
\begin{equation}\label{eq:gammabulk}
    \gamma_{bulk} = 12\int_0^R r^2dr\zeta \left(\frac{\partial \xi^r}{\partial r} + \frac{2}{r}\xi^r - l(l+1)\frac{\xi^{\perp}}{r}\right)^2 .
\end{equation}
where  $\zeta$ is the bulk viscous coefficient, $R$ is the radius of the star, $\xi^r$ and $\xi^{\perp}$ are radial and tangential component of the displacement vector respectively. \\
 
We compare this viscous dissipation timescale given by inverse of the rate $\tau_{bulk} = 1/\gamma_{bulk}$, with the inspiral timescale to confirm whether the hyperon bulk viscous dissipation can effectively drain energy from the decaying orbit during the inspiral. We consider leading order gravitational radiation where the GW emission takes energy from the orbit at a rate of
\begin{equation}\label{eq:GW_energy}
    \Dot{E}_{gw} = \frac{-32\mathcal{M}\Omega}{5c^5}(G\mathcal{M}\Omega)^{7/3}~.
\end{equation}
Here the chirp mass is given by $\mathcal{M} = M\left(\frac{q^3}{1+q}\right)^{1/5}$ with the primary star mass is $M$ and the companion mass is  $qM$. $\Omega $ denotes the orbital frequency which is given by $\Omega^2 = \frac{GM(1+q)}{D^3}$ where $D$ is the separation between the masses. In this evolution dynamics, the orbital timescale is given as~\cite{Andersson_2018}
\begin{equation}\label{eq:time}
     \frac{1}{t_D} \equiv \frac{\Dot{\Omega}}{\Omega} = \frac{96}{5c^5}(G\mathcal{M}\Omega)^{5/3}\Omega~.
\end{equation}
\begin{figure}
    \centering
    \includegraphics[width=0.5\textwidth]{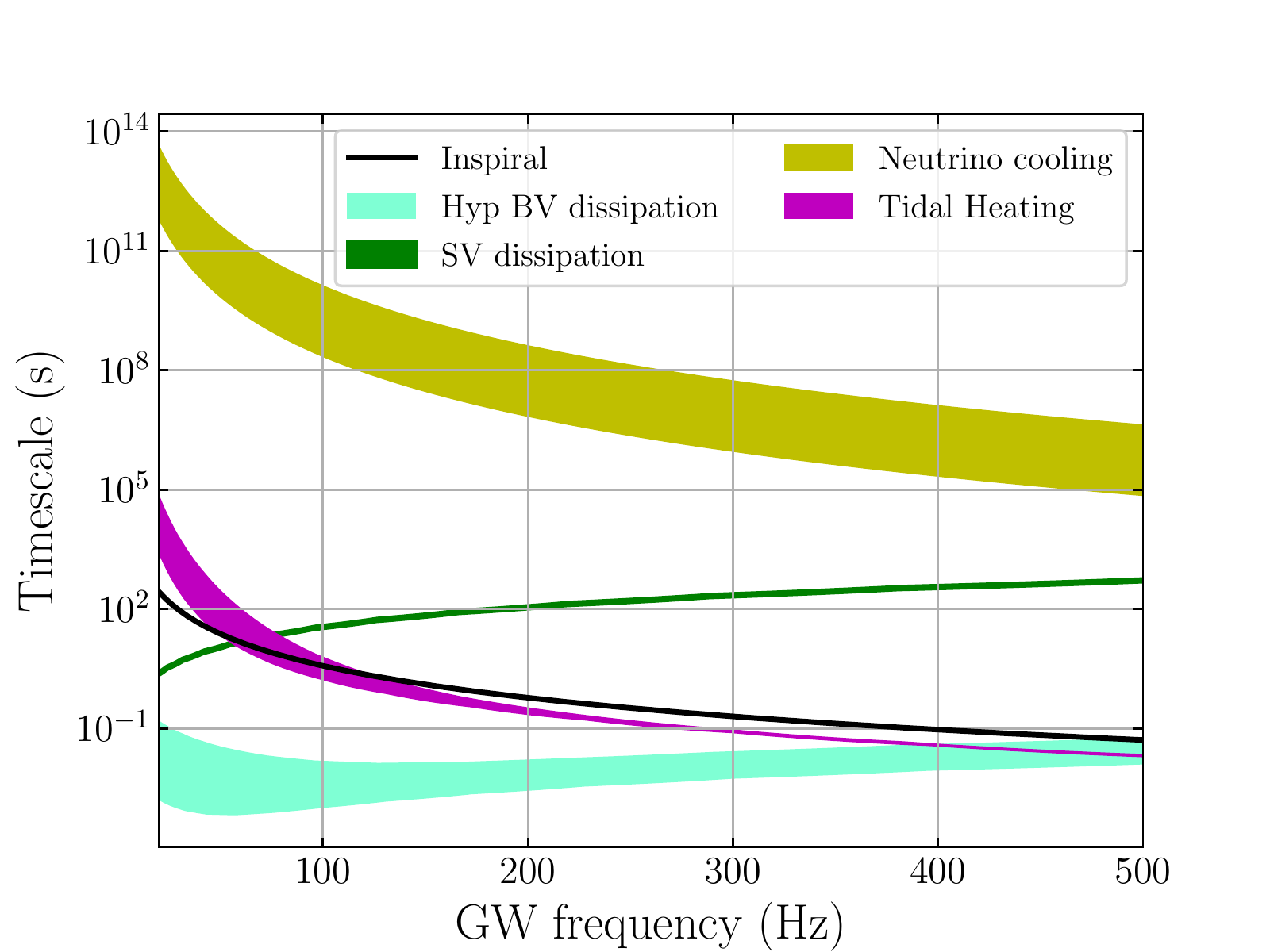}
    \caption{Estimated timescale for the different processes as a function of GW frequency compared against the inspiral timescale ($t_D$) for a NS binary of equal mass $1.8M_{\odot}$. Shear viscous and hyperon bulk viscosity dissipation corresponds to the dominant $f$-mode dissipation by shear viscosity from $e-e$ scattering~\cite{Lai_1994} and bulk viscosity from hyperons respectively. The tidal heating corresponds to the heating timescale from the hyperon bulk viscous dissipation. The different bands indicates uncertainities due to the choice of different EOSs given in Table~\ref{tab:table1}.}
    \label{fig:fig_2}
\end{figure}
In Fig.~\ref{fig:fig_2}, we plot this inspiral timescale for the LIGO-Virgo-KAGRA (LVK) frequency band of $20-500$ Hz~\citep{AdvLIGO2015,Abbott2016,AdvVIRGO2014,Kagra} with the viscous dissipation and tidal heating timescale for the case of equal binary mass of 1.8M$_{\odot}$ for all the EOSs given in Table.~\ref{tab:table1}. We see that both the hyperon bulk viscous dissipation and tidal heating timescales are smaller than the inspiral timescale, confirming that unlike shear viscous dissipation, hyperon bulk viscous dissipation and heating happens faster than the orbital evolution and can efficiently damp out the tidal energy  to heat up the star during the inspiral. \\

Since we only consider the dominant mode $l = m = 2$ $f$-mode contribution, the energy dissipation rate can be estimated from the mode amplitude~\cite{Lai_1994}
\begin{equation}\label{eq:viscener}
    \Dot{E}_{visc} = \frac{12\pi}{5}\frac{GM^2}{R}q^2(1+q)\omega_0^{-4}Q_0^2\left(\frac{R}{D}\right)^92\gamma_{bulk}
\end{equation}
where $Q_0$  tidal coupling strength of the $f$-mode and $\omega_0$ the normalised frequency of the $f$-mode. The heat content of the star due to the degenerate fermionic gas in the core can be given as $U \approx 4.5\times 10^{22}T^2 J$~\cite{Lai_1994}. During the inspiral, the timescales for cooling due to neutrino emission and surface photon luminosity are very high compared to the binary inspiral timescale as shown in Fig.~\ref{fig:fig_2}, and therefore negligible~\cite{Rees1992}. After integrating the thermal evolution equation $\frac{dU}{dt} = \Dot{E}_{visc} + \Dot{E}_{cool}$ from $D \to \infty$ when the stars were far apart and at a very low temperature ($10^5 - 10^6$K), we can get an estimate of the temperature reached as a function of their separation $D$
\begin{equation}\label{eq:final_T}
\begin{split}
    \left[\frac{T^4}{4} + Bln(T)\right] &=  \frac{\pi}{21870}\omega_0^{-4}Q_0^2q\frac{A}{10^{22}} \\ & \times \left(\frac{c^2R}{GM}\right)\left(\frac{c^3R^2}{G}\right)\left(\frac{3R}{D}\right)^5,
\end{split}    
\end{equation}
where $A$ and $B$ are parameters fitted to the functional dependence of $\gamma_{bulk}$ on the temperature ($T$), $\gamma_{bulk} = \frac{AT^2}{B+T^4}$ coming from the temperature dependence of timescale for hyperon bulk viscosity. In Table~\ref{tab:table1}, we provide the estimates of the temperature reached at a separation of $D = 3R$ when the stars are about to merge. We find temperatures $\sim 10^9 - 10^{10}$K, which are two orders of magnitude higher than the earlier estimates~\cite{Lai_1994,Arras2019}. 

\section{Phase error estimation and detectability}
\label{sec:phase}
\begin{figure}
    \centering
    \includegraphics[width=0.5\textwidth]{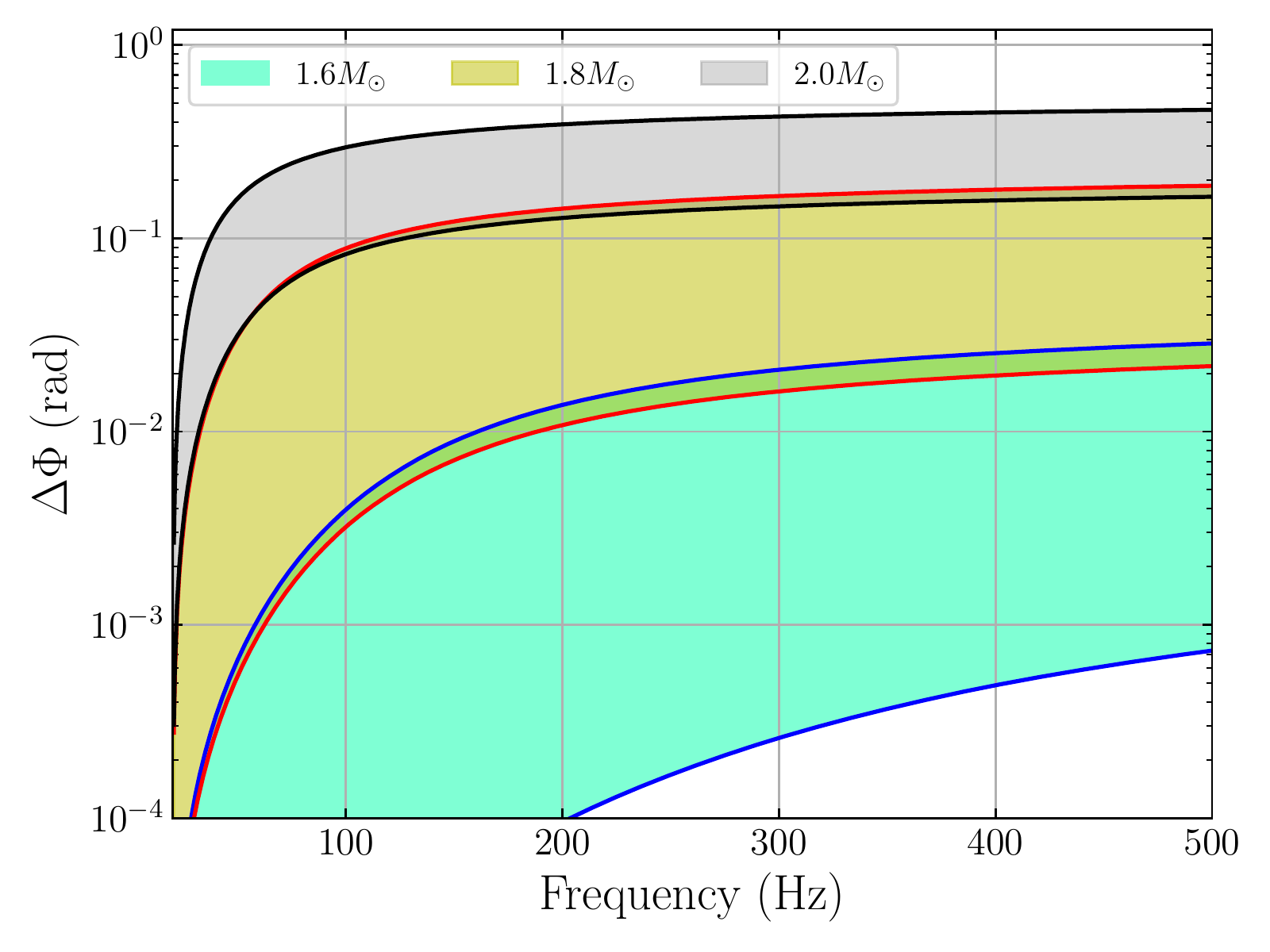}
    \caption{Estimated phase shift accumulated in the GW signal for equal mass binaries as a function of GW frequency. The different bands indicate uncertainties due to the choice of different EOSs given in Table~\ref{tab:table1} corresponding to different masses of equal mass binaries considered and the solid lines define their boundaries.}
    \label{fig:fig_3}
\end{figure}

The energy loss due to tidal heating during the binary inspiral will lead to a change in the number of wave cycles ($\Delta\mathcal{N}$) or equivalently, a phase shift $\Delta\Phi = 2\pi\Delta\mathcal{N}$ in the observed frequency range of the GW detectors. This is crucial to accurately guess the phase of the signal, otherwise the GW template can destroy a possible detection using matched filter technique~\cite{Matchfilter_2008}. This additional torque to the viscous dissipation of energy will lead to a total change in the number of cycles given by
\begin{equation}\label{eq:error}
   \Delta\mathcal{N}  = \frac{15\pi}{384}\left(\frac{c^2R}{GM}\right)^6 \omega_0^{-4}Q_0^2\frac{1}{q(1+q)}\left(\frac{R}{c}\right)^2\int_{f_a}^{f_b} \gamma_{bulk} df.    
\end{equation}
where $f_a$ is the frequency when the signal first enters the detector band, and $f_b$ , when it dives into the noise again. Since, $f$-modes are not resonantly excited, we need to do the integration over the whole frequency range unlike the cases for resonantly excited $g$-modes or $r$-modes where additional energy loss is associated with the particular mode frequency~\cite{Andersson_2018,Ma2021}. In Table~\ref{tab:table1}, we demonstrate the net phase difference accumulated at GW frequency of $500$ Hz for all the different EOSs and different values of equal mass binaries. In Fig.~\ref{fig:fig_3}, we display how this phase difference grows as a function of GW frequency, taking into consideration uncertainties due to the EOSs. For equal $1.6M_{\odot}$ binaries, we see that the net phase difference is of the order of $10^{-3} - 10^{-2}$ rad and for higher masses of $1.8M_{\odot}$ or $2M_{\odot}$, we get a net phase difference in the order of $0.1-0.5$ rad. \\

To be able to measure this phase difference using the current or future generation GW detectors, the phase uncertainty of detected GWs must be smaller than this extra phase shift. Recent analysis~\cite{Bruce_2021} have shown the phase error to be around $\Delta\phi \sim \pm 0.1$ rad for $f_{GW} \leq 300$ Hz inclusive of calibration uncertainties  for the GW170817 signal analysed using GW waveform model \textit{IMRPhenomPv2\_NRTidal}~\cite{NRTidal}. More recently, Read (2023)~\cite{Read_2023} compares a number of GW waveform models and shows that the uncertainty due to waveform differences is $\sim \pm 0.02$ rad for A+~\cite{ALIGO} and $\pm 10^{-3}$ rad for Cosmic Explorer (CE)~\cite{CE}. So, from these estimates, we see that a binary neutron star event with signal-to-noise ratio(SNR)  like GW170817 would produce enough tidal heating to be detectable using the current LVK detectors if it has a heavier component mass $\geq 1.8M_{\odot}$. In the 3G detectors, we can measure evidence of tidal heating due to hyperons even for much lower mass NS components. 

\section{Discussion}\label{sec:discussion}
In this work, we propose a novel effect of tidal heating in NSs during binary inspiral due to bulk viscosity originating from non-leptonic weak interaction processes involving hyperons in the NS core. We demonstrate the effect by calculating the bulk viscous dissipation of the dominant $f-$mode oscillations excited due to the tidal interaction during the inspiral. We consider several state-of-the-art EOSs including the hyperonic degrees of freedom consistent with multi-messenger observations. This dissipated energy can be effectively converted to thermal energy during the inspiral timescale and can heat up the stars upto $0.1 - 1$ MeV during the last orbits before coalescing. Although these estimates are higher by orders of magnitudes than earlier estimates by Lai~(1994)~\cite{Lai_1994} and Arras et al.~(2019)~\cite{Arras2019}, they are not sufficiently high to demand inclusion of thermal effects which are very relevant for post-merger studies~\cite{Fields_2023} ,in the EOS during inspiral~\cite{Thermal_2022}. The induced phase shift in the GW signal due to this tidal heating is $\sim 0.1$ rad for component masses $\geq 1.8M_{\odot}$ which is potentially detectable by current and future generation GW detectors.
\\

This work suggests for the first time that tidal heating can be used as a probe of strange matter inside NS core, and opens up the possibility of exploring the out-of-equilibrium effects of NS matter on the GW signals. These findings pave the way for more sophisticated analyses as followup of this work, including a consistent general relativistic formulation of tidal heating and f-mode calculations, other EOS models including the contribution from other species of hyperons, other forms of strange matter (e.g. quarks) in order to get a more quantitatively accurate estimate of the observable effects. There are various other directions in which our work can be further developed. First, a detailed investigation of the Post-Newtonian(PN) order at which the tidal heating becomes relevant will have to be performed to confirm the conclusions related to the qualitative Newtonian inspiral discussed in this work. Second, resonant $g$-mode contribution to tidal heating can also be as significant~\cite{Lai_1994} since the hyperon bulk visocity value is one order of magnitude order higher than those of $f-$modes (see Fig.~\ref{fig:fig_1}). Third,  bulk viscosity from the dominant channel in unpaired strange quark matter is $\sim 10^{29}-10^{30}$ gm cm$^{-1}$ s$^{-1}$ around $10^8-10^9$K~\cite{Alford_2010,Alford_2014} and could also contribute to the tidal heating. Finally, the effect of superfluidity on the hyperon bulk viscosity should be investigated since they are known to reduce the rates of the weak interactions and also effective below the critical temperature of $10^9$K~\cite{Haskell2010}. Several of these followup studies are currently in progress and will be reported in forthcoming publications.\\

We are currently developing a detailed Bayesian analysis using post-Newtonian waveform including the effect of viscous dissipation to accurately determine the bulk viscosity as well as estimate any possible degeneracy with other intrinsic parameters of binary NS inspiral. Considering the recent order-of-magnitude post-Newtonian estimate~\cite{Most_2021} of the direct effect of bulk viscosity on the orbital motion  at $f = 100$Hz, when the bulk viscosity is $\sim 10^{32}$ gm cm$^{-1}$ s$^{-1}$(from Fig.~\ref{fig:fig_1}), one would require  SNR of $100$ to get a $10\%$ measurement on the bulk viscosity, which is achievable using the third generation GW detectors. Recent observations of binary neutron star merger  GW190425~\cite{GW190425} or neutron star-black hole merger GW200105~\cite{NSBh} having component NS masses $m_1^{90\%} = [1.6,1.87]M_{\odot}$ and $m_2 = 1.9^{+0.3}_{-0.2}$ respectively, suggest that there might be other formation channels for higher mass NS in binary systems which can have sufficient hyperons in their core to contribute significantly to the tidal heating that can be detectable.  A future detection of this high bulk viscous effect originating from non-leptonic weak interaction processes involving hyperons either in terms of deviation in the orbital motion or tidal heating would strongly indicate the presence of strange hyperons in the NS interior, having a great significance in nuclear astrophysics. Even a non-detection of this effect can place an upper limit on the bulk viscosity and the hyperon fraction in the NS core, with important implications for understanding of dense nuclear matter. \\

\begin{acknowledgments}
 The authors thank Alexander Haber for his useful comments regarding the paper. The authors would like to acknowledge the usage of the high-performance supercomputer Pegasus at Inter-University Centre for Astronomy and Astrophysics for the numerical calculations.
\end{acknowledgments}

\nocite{*}

\bibliography{apssamp}

\end{document}